\documentclass[cameraready]{Interspeech}

\usepackage{hyperref}
\usepackage{xcolor}
\usepackage{booktabs}
\usepackage{multirow}
\usepackage{adjustbox}
\usepackage[table,dvipsnames]{xcolor}
\usepackage[table]{xcolor}

\usepackage{verbatim}
\usepackage{amssymb}

\usepackage{subcaption}
\usepackage{url}
\usepackage{graphicx} 
\usepackage{pifont}

\definecolor{sb}{RGB}{0,180,180} 
\definecolor{fern}{RGB}{113,159,117}
\definecolor{rosewood}{RGB}{170,109,118}   

\usepackage{amsmath} 

\makeatletter
\let\tipa@bang\!
\let\tipa@semicolon\;
\let\tipa@dblbar\|

\DeclareRobustCommand{\!}{\ifmmode\mskip-\thinmuskip\else\tipa@bang\fi}
\DeclareRobustCommand{\;}{\ifmmode\mskip\thickmuskip\else\tipa@semicolon\fi}
\DeclareRobustCommand{\|}{\ifmmode\Vert\else\tipa@dblbar\fi}
\makeatother

\title{Synergizing Zero-Shot Cross-Lingual Alzheimer Detection with Language-Invariant Multimodal Bi-Geometric Adversarial Learning}

\author[affiliation={1}, orcid=0009-0004-2926-7777, equalcontribution]{Girish}{}
\author[affiliation={1}, orcid=0009-0000-1982-7110, equalcontribution]{Mohd Mujtaba}{Akhtar}
\author[affiliation={2}, orcid=0009-0009-9371-6983, equalcontribution]{Farhan}{Sheth}
\author[affiliation={1}, orcid=0009-0008-2638-7000, correspondingauthor]{Muskaan}{Singh}
\author[affiliation={1}, orcid=0000-0003-3985-4022]{Juliana}{Gerard}
\author[affiliation={1}, orcid=0000-0002-3906-1172]{Paula}{McClean}
\author[affiliation={1}, orcid=0000-0001-8724-4398]{Kongfatt}{Wong-Lin}

\address{
$^1$ Ulster University, UK,
$^2$ Manipal University, India
}

\email{
girish.research.pr@gmail.com,
mmakhtar.research@gmail.com,
m.singh@ulster.ac.uk
}

\keywords{Dementia detection, Multimodal learning, Cross-lingual, Multimodal fusion, Zero-shot learning}

\usepackage{comment}

\begin{document}

\maketitle


\begin{abstract}
In this work, we study zero-shot cross-lingual speech-based Alzheimer’s disease detection (SADD). We hypothesize that learning language-invariant multimodal representations by fusing multilingual speech and text pretrained models is essential for reliable transfer to unseen languages, as the two modalities capture complementary acoustic and linguistic markers of cognitive impairment while adversarial learning suppresses language-specific confounds. Empirical results in zero-shot cross-lingual evaluation substantiate the hypothesis, showing that multimodal fusion consistently outperforms unimodal baselines. To this end, we propose \textbf{\texttt{ORBIT}}, a novel framework that combines cross-attentive fusion, multi-tap language adversaries, and complementary spherical–hyperbolic geometric learning with consensus clustering. Across settings, \textbf{\texttt{ORBIT}} achieves the strongest performance compared to unimodal models and simple concatenation-based fusion baselines.


\end{abstract}

\section{Introduction}

Speech-based Alzheimer’s disease detection (SADD) plays a pivotal role in scalable, non-invasive cognitive screening, enabling computational systems to identify subtle linguistic and acoustic signatures of neurodegeneration \cite{luz20_interspeech,luz21_interspeech,karlekar2018detecting}. Spontaneous speech offers rich acoustic–linguistic evidence of decline, from lexical–syntactic simplification to disfluency and prosodic change, enabling effective AD screening. SADD has far-reaching implications, from enabling low-burden remote screening and longitudinal monitoring in telehealth to improving cohort enrichment and outcome tracking in clinical trials~\cite{hakkanitur10_interspeech,GOLD2018234}. Early research largely relied on hand-crafted acoustic and linguistic descriptors—e.g., cepstral/prosodic statistics and pause- and fluency-based measures—paired with classical machine-learning classifiers such as SVM and random forests~\cite{gosztolya16_interspeech,weiner18_interspeech,nasreen21_interspeech}. A major shift followed with deep learning architectures—ranging from convolutional models on acoustic representations to recurrent sequence models and multimodal LSTM-based fusion—substantially improving SADD from spontaneous speech~\cite{warnita18_interspeech,rohanian20_interspeech,rohanian21_interspeech}. By the end of the last decade, the landscape of SADD had shifted markedly with the rise and wide-scale availability of self-supervised learning (SSL) and large pretrained models (PTMs). Trained on massive and diverse corpora, these PTMs provide strong, transferable representations that substantially boost performance and reduce the need to train task-specific models from scratch on limited clinical data. Collectively, these PTMs have catalyzed rapid development in SADD. As such, researchers have explored a broad range of state-of-the-art speech and language PTMs (e.g., wav2vec 2.0–style SSL encoders and BERT-like text encoders) for SADD. Zhu et al. \cite{zhu21e_interspeech} proposed WavBERT, integrating wav2vec-based ASR representations with BERT to leverage both semantic and non-semantic cues. Balagopalan et al. Gauder et al. \cite{gauder21_interspeech} evaluated speech embeddings from multiple PTMs (including wav2vec 2.0) for Alzheimer’s recognition, further supporting the effectiveness of PTM-based representations. These PTMs predominantly use attention-based architectures to capture discourse- and prosody-level context present in spontaneous clinical speech. In recent years, a new class of multilingual speech foundation models (SFMs) has captured significant attention in the community—most notably Whisper-style encoder–decoder models and SSL multilingual encoders such as mHuBERT~\cite{radford2023robust,boito2024mhubert}. These PTMs are pretrained on large, diverse speech corpora and yield highly transferable representations that are well-suited to the variability of spontaneous clinical speech, particularly when labeled data are limited. Prior research has also demonstrated the utility of mHuBERT embeddings for speech-based cognitive screening~\cite{azadmaleki2025speechcare}. \par
In this study, we move beyond language-specific evaluation and target zero-shot cross-lingual SADD, and \textit{hypothesize that reliable transfer hinges on learning language-invariant multimodal representations: fusing speech and text PTMs preserves complementary acoustic--linguistic markers of impairment, while adversarial objectives suppress language-specific confounds and spurious cues that would otherwise hinder generalization}. To the best of our knowledge, this work is the first to study multimodal PTM fusion for SADD under explicit language-invariance constraints, with zero-shot cross-lingual evaluation protocols. To our end, we propose, \textbf{\texttt{ORBIT}} – \textbf{O}ptimized \textbf{R}epresentation Learning via \textbf{BI}-geometric and Adversarial \textbf{T}raining for Zero-shot Cross-lingual Alzheimer Detection, a novel framework for language-invariant cross-lingual SADD that fuses multilingual speech and text PTMs. It performs bidirectional cross-attention to align the audio and text streams and integrates them into a unified representation that captures complementary acoustic–linguistic markers of impairment. To prevent the fused space from encoding language identity, \textbf{\texttt{ORBIT}} introduces multi-tap adversarial learning, applying gradient-reversal discriminators at the fusion level, after each geometric projection head, and at the cluster-assignment level to suppress residual language cues. In parallel, \textbf{\texttt{ORBIT}} projects the fused representation into complementary spherical and hyperbolic manifolds, capturing distinct structural aspects of the impairment signal, and refines sample organization via consensus clustering across the two geometries. Finally, \textbf{\texttt{ORBIT}} performs prototype-based classification with product-of-experts voting, producing stable decisions that generalize under cross-lingual distribution shift. \newline 
\noindent \textbf{The key contributions of our study are as follows:} 
(i) We propose \textbf{\texttt{ORBIT}}, a language-invariant framework for zero-shot cross-lingual SADD that fuses multilingual speech and text PTMs and reduces language leakage via multi-tap adversarial learning at the fusion, geometric, and clustering levels. 
(ii) We curate a multilingual SADD benchmark (English, Spanish, Chinese, Greek) and evaluate under LOLO and LTLO, demonstrating consistent gains over unimodal/multimodal PTM baselines. \newline
\footnote{Project resources are publicly available at: \url{https://github.com/Helixometry/ORBIT.git}.}

\vspace{-0.32cm}
\section{Representations}
\noindent \textbf{Audio Foundation Models:} mHuBERT-14 \cite{boito2024mhubert}, a 95M-parameter multilingual HuBERT variant trained on $\sim$90k hours of open-licensed speech covering 147 languages. Whisper (base)\cite{radford2023robust}, a 74M-parameter encoder–decoder Transformer trained on $\sim$680k hours of weakly supervised, diverse audio. wav2vec 2.0 (base)\cite{baevski2020wav2vec}, which learns contextualized representations by masking latent speech features and optimizing a contrastive objective against quantized targets, enabling strong ASR with limited labeled data. MMS-1B\cite{pratap2023mms}, a 1B-parameter model pretrained with the wav2vec 2.0 objective on $\sim$500k hours spanning over 1{,}400 languages, XLS-R-1B\cite{babu2021xls}, a 1B-parameter cross-lingual model trained on 436k hours of unlabeled speech in 128 languages. For all experiments, we preprocess audio by resampling to 16kHz prior to feature extraction. To obtain fixed-length utterance representations, we mean-pool the final hidden states over time. The resulting embedding dimensionalities are: 768 (mHuBERT-147), 512 (Whisper-base), 768 (wav2vec 2.0 base), and 1280 for both MMS-1B and XLS-R-1B. \par
\noindent \textbf{Text Foundation Models:} BERT \cite{devlin2019bert}, a bidirectional Transformer pre-trained with masked-language modeling and next-sentence prediction on large text corpora. We use the bert-base-multilingual-cased checkpoint. XLM-RoBERTa\cite{conneau2019unsupervised}, a multilingual masked-language model trained on approximately 2.5 TB of filtered CommonCrawl spanning 100 languages; we use the xlm-roberta-large variant. E5-large\cite{wang2024multilingual} is a multilingual embedding model trained on $\sim$1B cross-lingual text pairs via contrastive pre-training and fine-tuning. Qwen-3-Embeddings\cite{qwen3embedding} is a multilingual embedding and re-ranking model derived from the Qwen3 family, trained with large-scale synthetic data, supervised fine-tuning, and model merging, and reports competitive performance on MTEB and related benchmarks. For BERT, we derive sentence vectors by applying attention-mask-aware mean pooling to the final hidden layer. For XLM-RoBERTa, we compute the mean of the final hidden states across the token dimension. For the embedding-specialized models (E5-large and Qwen-3-Embeddings), we utilize the models’ output embeddings. Inputs are tokenized with each model’s native tokenizer. The resulting embedding dimensionalities are 768 for BERT, and 1024 for XLM-RoBERTa, E5-large, and Qwen-3-Embeddings.

\section{Modeling}


\noindent \textbf{Downstream Modeling:} We evaluate each audio and text encoder with two lightweight classifiers: an FCN and a 1D-CNN. The CNN uses two convolution layers (64 and 128 filters, kernel 3) with ReLU and max-pooling, followed by a 128-unit dense layer and a final softmax classifier. The FCN simply flattens the input features and applies a 256-unit ReLU layer before the softmax output.

\begin{figure*}[!htb]
    \centering
    \includegraphics[width=0.6\linewidth]{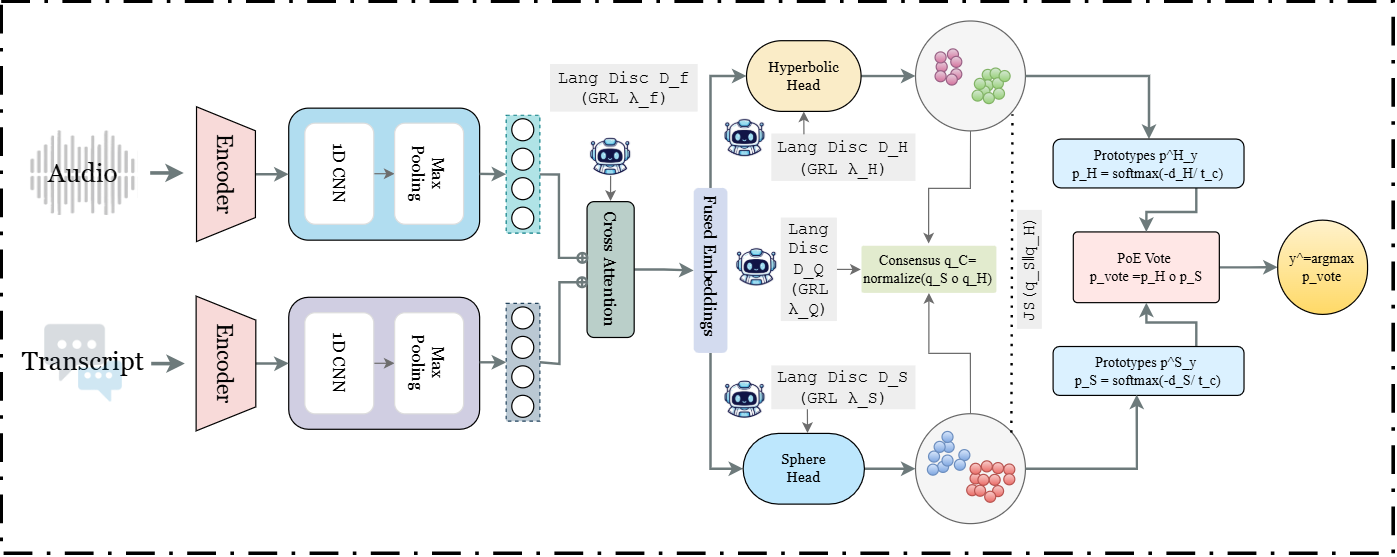}
    \caption{Proposed Framework: \textbf{\texttt{ORBIT}}}
    \label{archi}
\vspace{-15pt}
\end{figure*}

\subsection{Proposed Framework: \textbf{\texttt{ORBIT}}}
We propose \textbf{\texttt{ORBIT}}, a multimodal framework for SADD that enables zero-shot cross-lingual detection through language-invariant fusion and structured geometric learning. The overall architecture is illustrated in Figure~\ref{archi}. \newline
We denote each input as $x=(x_a,x_t)$, where $x_a$ and $x_t$ are audio and text features.  Each sample has a clinical label $y \in \{0,1\}$ and a language label $\ell \in \{1,\dots,|\mathcal{L}|\}$, used only in training. Bold lowercase letters represent vectors, and bold uppercase letters represent sequences. $\|\cdot\|$ denotes the $\ell_2$ norm, $\mathrm{CE}$ and $\mathrm{KL}$ denote cross-entropy and Kullback–Leibler divergence, and $[\cdot \| \cdot]$ denotes concatenation.  \newline
\noindent \textbf{Encoders and pooling:} We encode the audio and transcript with $\phi_a$ and $\phi_t$ to obtain sequences $\mathbf{S}=[\mathbf{s}_1,\ldots,\mathbf{s}_{T_a}]$ and $\mathbf{H}=[\mathbf{h}_1,\ldots,\mathbf{h}_L]$, optionally refined by lightweight temporal 1D-convs, and then apply attention pooling to produce fixed-length embeddings $\mathbf{a}$ and $\mathbf{t}$:
{\scriptsize \[ \begin{aligned} \alpha_i &= \frac{\exp(\bm{v}^{\!\top}\tanh(W\,\mathbf{s}_i))}{\sum_{j}\exp(\bm{v}^{\!\top}\tanh(W\,\mathbf{s}_j))}, &\mathbf{a} &= \sum_i \alpha_i\,\mathbf{s}_i,\\ \beta_j &= \frac{\exp(\bm{u}^{\!\top}\tanh(U\,\mathbf{h}_j))}{\sum_{k}\exp(\bm{u}^{\!\top}\tanh(U\,\mathbf{h}_k))}, &\mathbf{t} &= \sum_j \beta_j\,\mathbf{h}_j \end{aligned} \] }

\noindent Next, we perform bidirectional cross-attention to condition each modality on the other (i.e., using one embedding as the query and the other as context), and fuse the resulting representations with an MLP layer.

\noindent \textbf{Language adversaries and bi-geometric projections:} To encourage language-invariant fusion, we attach a language discriminator $D_f$ to the fused representation $\mathbf{f}$ through a gradient-reversal layer (GRL), so that $\mathbf{f}$ remains predictive for the clinical label while being uninformative about the language label $\ell$:
{\scriptsize
\[
\mathcal{L}_{\mathrm{adv},f}=\mathrm{CE}\!\big(D_f(\mathrm{GRL}(\mathbf{f})),\ell\big)
\]
}

\noindent In parallel, we project $\mathbf{f}$ into two complementary geometric spaces to capture different structure in the impairment signal. On the sphere $\mathbb{S}^{D^S}(r)$, we normalize the projected features and compare them using spherical geodesic distance:
{\scriptsize
\[
\mathbf{x}^S=\frac{r\,W^S\mathbf{f}}{\|W^S\mathbf{f}\|},\qquad
d_S(\mathbf{x},\mathbf{y})=r\,\arccos\!\Big(\frac{\langle\mathbf{x},\mathbf{y}\rangle}{r^2}\Big)
\]
}

\noindent We additionally map $\mathbf{f}$ to the Poincaré ball $\mathbb{B}^{D^H}_c$ to model hierarchical or branching structure, and measure similarity via hyperbolic geodesic distance:
{\scriptsize
\[
\begin{aligned}
\mathbf{x}^H &= \tanh\!\big(\sqrt{c}\,\|W^H\mathbf{f}\|\big)\,
               \frac{W^H\mathbf{f}}{\sqrt{c}\,\|W^H\mathbf{f}\|},\\
d_H(\mathbf{x},\mathbf{y}) &= \frac{1}{\sqrt{c}}\cosh^{-1}\!\left(
1+\frac{2c\|\mathbf{x}-\mathbf{y}\|^2}{(1-c\|\mathbf{x}\|^2)(1-c\|\mathbf{y}\|^2)}
\right)
\end{aligned}
\]
}

\noindent Since language cues can re-emerge after the non-linear geometric projections, we attach GRL-based language discriminators to both geometric embeddings, $\mathbf{x}^S$ and $\mathbf{x}^H$, to further suppress language information:
{\scriptsize
\[
\resizebox{\linewidth}{!}{$
\mathcal{L}_{\mathrm{adv},S}=\mathrm{CE}\!\big(D_S(\mathrm{GRL}(\mathbf{x}^S)),\ell\big),\;
\mathcal{L}_{\mathrm{adv},H}=\mathrm{CE}\!\big(D_H(\mathrm{GRL}(\mathbf{x}^H)),\ell\big)
$}
\]
}

\noindent To model intra-class heterogeneity, we learn $K{>}2$ centers in each manifold and compute temperature-scaled soft assignments based on the corresponding geodesic distances:
{\scriptsize
\[
\resizebox{\linewidth}{!}{$
q_S(k)=\frac{\exp(-d_S(\mathbf{x}^S,\hat{\mathbf{c}}^S_k)/\tau)}{\sum_j \exp(-d_S(\mathbf{x}^S,\hat{\mathbf{c}}^S_j)/\tau)},\;
q_H(k)=\frac{\exp(-d_H(\mathbf{x}^H,\hat{\mathbf{c}}^H_k)/\tau)}{\sum_j \exp(-d_H(\mathbf{x}^H,\hat{\mathbf{c}}^H_j)/\tau)}
$}
\]
}

\noindent We then combine the two views via a product-of-experts (PoE) consensus, which emphasizes clusters supported by both geometries:
{\scriptsize
\[
q_C(k)=\frac{q_S(k)\,q_H(k)}{\sum_j q_S(j)\,q_H(j)}
\]
}

\noindent Finally, we regularize the assignments with Jensen--Shannon agreement between $q_S$ and $q_H$, DEC-style sharpening toward $p_{\mathrm{DEC}}(q_C)$, and a prototype margin term (defined below). To prevent residual language coding at the assignment level, we add a cluster-level adversary on the concatenated responsibilities:
{\scriptsize
\[
\mathcal{L}_{\mathrm{adv},Q}=\mathrm{CE}\!\big(D_Q(\mathrm{GRL}([q_S\|q_H])),\ell\big)
\]
}

\noindent Next, we represent each class $y\in\{0,1\}$ with prototypes $(\mathbf{p}^S_y,\mathbf{p}^H_y)$ in the spherical and hyperbolic spaces, and convert prototype distances into temperature-scaled posteriors:
{\scriptsize
\[
\resizebox{\linewidth}{!}{$
p_S(y)=\frac{\exp\!\big(-d_S(\mathbf{x}^S,\mathbf{p}^S_y)/\tau_c\big)}
{\sum_{y'} \exp\!\big(-d_S(\mathbf{x}^S,\mathbf{p}^S_{y'})/\tau_c)},\;
p_H(y)=\frac{\exp\!\big(-d_H(\mathbf{x}^H,\mathbf{p}^H_y)/\tau_c\big)}
{\sum_{y'} \exp\!\big(-d_H(\mathbf{x}^H,\mathbf{p}^H_{y'})/\tau_c)}
$}
\]
}

\noindent We combine the two geometric views using a product-of-experts (PoE) vote, which favors labels supported by both heads:
{\scriptsize
\[
p_{\text{vote}}(y)=\frac{p_S(y)\,p_H(y)}{\sum_{y'} p_S(y')\,p_H(y')}
\]
}

\noindent Finally, \textbf{\texttt{ORBIT}} is trained end-to-end with a clinical classification loss together with clustering-consensus regularizers and multi-tap adversarial objectives. Let $\mathcal{Z}=\{\mathbf{f},\mathbf{x}^S,\mathbf{x}^H,[q_S\|q_H]\}$ denote the tapped representations and $\mathcal{D}=\{D_f,D_S,D_H,D_Q\}$ the corresponding language discriminators. The resulting optimization can be written as:
{\scriptsize
\[
\begin{aligned}
\min_{\Theta}\ \mathcal{L}_{\mathrm{cls}}
&+ \lambda_{\mathrm{BGCC}}\!\left(\mathcal{L}_{\mathrm{dec}}+\mathcal{L}_{\mathrm{js}}+\mathcal{L}_{\mathrm{margin}}\right)\\
&+ \sum_{Z\in\mathcal{Z}} \lambda_Z \max_{D_Z}\ \mathbb{E}\!\left[\mathrm{CE}(D_Z(Z),\ell)\right]
\end{aligned}
\]
}

\noindent We implement the min--max terms using GRLs for stable training. This multi-tap design is important because language information can re-enter at different stages; applying adversarial constraints at $\{\mathbf{f},\mathbf{x}^S,\mathbf{x}^H,[q_S\|q_H]\}$ progressively removes such leakage wherever it arises. In parallel, the spherical and hyperbolic heads provide complementary inductive biases, while PoE-based consensus and DEC/JS refinement stabilize the cluster structure and reduce language-coded groupings. Together, these components promote decision boundaries driven by impairment cues rather than language identity, enabling reliable zero-shot cross-lingual performance. We train for 50 epochs (batch size 32) with AdamW using dropout and early stopping. Encoders are frozen for 2--3 epochs, then the top $L$ layers are unfrozen with a smaller LR than the heads. The number of trainable parameters in \textbf{\texttt{ORBIT}} ranges from 4.2M to 5.5M, depending on the encoder dimensionality.

\section{Experiments}

\subsection{Dataset} 

We construct a multilingual corpus covering four languages (English, Chinese, Spanish, Greek) for ADD using publicly available resources. The corpus consists of:  
\textbf{(i) Pitt} \cite{becker1994natural}: English recordings and transcripts from the longitudinal Pittsburgh DementiaBank corpus, we use Cookie Theft picture description task (Samples: HC=243, AD=309).
\textbf{(ii) Ivanova} \cite{ivanova2022discriminating}: Spanish standardized reading samples from 361 older adults, collected under controlled acoustic conditions (Samples: HC=196, AD=74).  
\textbf{(iii) NCMMSC} \cite{ncmmsc2021ad}: Chinese recordings from the National Conference on Man–Machine Speech Communication, including picture description and fluency tasks (Samples: HC=108, AD=79). 
\textbf{(iv) Dem@Care} \cite{karakostas2016care}: Greek recordings collected in home-like environments; we use subsets DS3, DS5, and DS7 (Samples: HC=58, AD=115). Since official transcripts are unavailable, we generate them using Whisper-large-v3. To enable a unified zero-shot cross-lingual protocol across the heterogeneous corpora considered in this work--whose label sets and class prevalences are not fully aligned—we primarily cast SADD as a binary task (AD vs. HC) with consistent class balancing and evaluation across languages.

\subsection{Experimental Results}
Table~\ref{tab:1} reports results for individual PTMs on the combined multilingual set under two downstream architectures. CNN models generally outperform FCN across most audio and text PTMs on the combined multilingual set. FCN generally performs competitively but remains below CNN, indicating that the choice of neural downstream head can substantially influence the effectiveness of individual PTMs. Across modalities, mHuBERT (mH) is the strongest audio backbone, while BERT(BT) consistently leads among text encoders under both FCN and CNN. Its stronger performance likely reflects the richer multilingual pretraining and higher-capacity representations, which better capture clinically relevant acoustic and linguistic patterns in spontaneous speech. Also, we observe that PTM performance varies across downstream architectures. Such sensitivity to the downstream architecture has also been noted in prior work on pretrained speech representations~\cite{yang21c_interspeech}. Within the unimodal baselines, mHuBERT is the strongest audio PTM, with Whisper and wav2vec2 performing competitively but below it, while MMS and XLS-R are comparatively weaker on this pooled setting. On the text side, BERT leads, followed by E5 and Qwen3, whereas XLM-R reports the lowest scores, indicating that not all multilingual encoders transfer equally well to SADD. These unimodal scores serve as our reference baselines for the subsequent cross-lingual fusion experiments. 
\begin{table}[!hbt]
\centering
\scriptsize
\setlength{\tabcolsep}{18pt}
\begin{adjustbox}{width=\linewidth,center}
\begin{tabular}{l|cc|cc}
\toprule
\multirow{3}{*}{\textbf{PTMs}} & \multicolumn{4}{c}{\textbf{E+C+S+G}} \\ 
\cmidrule(lr){2-5}
& \multicolumn{2}{c|}{\textbf{FCN}} & \multicolumn{2}{c}{\textbf{CNN}} \\ 
\cmidrule(lr){2-3} \cmidrule(lr){4-5}
& \textbf{A} & \textbf{F1} & \textbf{A} & \textbf{F1} \\ 
\midrule
\multicolumn{5}{c}{\textbf{AUDIO}} \\ 
\midrule
\textbf{mH}  & \cellcolor{sb!90} 93.46 & \cellcolor{sb!90} 92.69 & \cellcolor{sb!90} 95.66 & \cellcolor{sb!90} 94.05 \\
\textbf{WP}  & \cellcolor{sb!30} 87.77 & \cellcolor{sb!30} 82.83 & \cellcolor{sb!56} 93.41 & \cellcolor{sb!56} 91.88 \\
\textbf{WV2} & \cellcolor{sb!56} 89.10 & \cellcolor{sb!56} 84.08 & \cellcolor{sb!30} 90.44 & \cellcolor{sb!30} 87.99 \\
\textbf{MS}  & 81.94 & 80.34 & 85.69 & 83.19 \\
\textbf{XS}  & 83.91 & 81.45 & 88.62 & 86.01 \\ 
\midrule
\multicolumn{5}{c}{\textbf{TEXT}} \\ 
\midrule
\textbf{BT}  & \cellcolor{sb!90} 91.14 & \cellcolor{sb!90} 89.06 & \cellcolor{sb!90} 94.53 & \cellcolor{sb!90} 92.61 \\
\textbf{XL}  & 83.64 & 81.19 & 85.48 & 84.12 \\
\textbf{E5}  & \cellcolor{sb!56} 87.31 & \cellcolor{sb!56} 85.29 & \cellcolor{sb!30} 89.25 & \cellcolor{sb!30} 87.95 \\
\textbf{Q3}  & \cellcolor{sb!30} 86.23 & \cellcolor{sb!30} 84.63 & \cellcolor{sb!56} 90.77 & \cellcolor{sb!56} 88.02 \\ 
\bottomrule
\end{tabular}
\end{adjustbox}
\caption{Performance of audio and text pretrained models on the combined multilingual set (E+C+S+G), reported as Accuracy (A) and macro F1 using FCN and CNN downstream classifiers. Abbreviations: mH=mHuBERT-147, WP=Whisper-base, WV2=wav2vec,2.0, MS=MMS-1B, XS=XLS-R-1B, BT=BERT, XL=XLM-R, E5=E5-large, Q3=Qwen-3-Embeddings. Abbreviations are applied consistently in Table~\ref{tab:2}.}
\label{tab:1}
\vspace{-30pt}
\end{table}
Table~\ref{tab:2} present the result for cross-lingual performance under leave-two-languages-out (LTLO) and leave-one-language-out (LOLO) zero-shot protocols, comparing standard baselines against the proposed \textbf{\texttt{ORBIT}} framework. Following Table~\ref{tab:1}, we adopt the CNN head for all subsequent experiments due to its consistently stronger performance. In LTLO, models are trained on English and Chinese and evaluated on the two held-out languages, with scores averaged across both test languages. In LOLO, models are trained on three languages and evaluated on the remaining one; we report mean performance over the two LOLO runs, respectively. As expected, LOLO yields higher performance than LTLO due to broader training coverage. We then evaluate both the baseline setup and \textbf{\texttt{ORBIT}} in a staged manner: (i) audio-only and (ii) text-only settings, followed by (iii) audio+text fusion. For fusion, we use simple concatenation as a strong baseline and compare it against \textbf{\texttt{ORBIT}} with and without cross-attention under the same training protocol. Our findings reveal that \textbf{\texttt{ORBIT}} consistently outperforms both individual modalities and the concatenation-based fusion baseline under LTLO and LOLO, underscoring the value of language-invariant cross-modal alignment for zero-shot cross-lingual SADD. When examining audio+text fusion, we observe that simple concatenation already improves over unimodal models in many cases, but enabling cross-attention within \textbf{\texttt{ORBIT}} yields further gains, indicating that explicitly aligning the two modalities is more effective than treating them as independent feature blocks. 
With \textbf{\texttt{ORBIT}}, we also observe stronger complementary behavior between speech and text PTMs than what is achieved by simple concatenation. In particular, \textbf{\texttt{ORBIT}} with cross-attention achieves the best LOLO performance with the mHuBERT + Qwen3 pairing (86.98 accuracy / 85.29 macro-F1), highlighting that carefully aligned multimodal fusion can unlock gains beyond what either modality provides alone. Under the LTLO setting, the strongest performance is obtained with mHuBERT + E5 (85.49), while the best macro-F1 is reached with Whisper + E5 (83.34), suggesting that the most effective pairing can depend on whether the objective prioritizes accuracy or balanced class-wise performance.
\begin{table}[!hbt]
\centering
\scriptsize
\begin{adjustbox}{width=\linewidth,center}
\begin{tabular}{l|cccc|cccc}
\toprule
\multirow{3}{*}{\textbf{PTMs}} 
 & \multicolumn{4}{c}{\textbf{BASELINE}} 
 & \multicolumn{4}{c}{\textbf{ORBIT}} \\ 
\cmidrule(lr){2-5} \cmidrule(lr){6-9}
 & \multicolumn{2}{c}{\textbf{LTLO}} & \multicolumn{2}{c}{\textbf{LOLO}} 
 & \multicolumn{2}{c}{\textbf{LTLO}} & \multicolumn{2}{c}{\textbf{LOLO}} \\ 
\cmidrule(lr){2-3} \cmidrule(lr){4-5} \cmidrule(lr){6-7} \cmidrule(lr){8-9}
 & \textbf{A} & \textbf{F1} & \textbf{A} & \textbf{F1} 
 & \textbf{A} & \textbf{F1} & \textbf{A} & \textbf{F1} \\ 
\midrule
\multicolumn{9}{c}{\textbf{AUDIO}} \\ 
\midrule
\textbf{mH}  & \cellcolor{sb!56} 61.38 & \cellcolor{sb!56} 60.11 & \cellcolor{sb!90} \textbf{66.40} & \cellcolor{sb!56} 62.01 & \cellcolor{sb!56} 73.32 & \cellcolor{sb!56} 71.61 & \cellcolor{sb!56} 78.93 & \cellcolor{sb!56} 75.20 \\
\textbf{WP}  & \cellcolor{sb!90} \textbf{62.55} & \cellcolor{sb!90} \textbf{61.05} & \cellcolor{sb!56} 65.33 & \cellcolor{sb!90} \textbf{63.18} & \cellcolor{sb!90} \textbf{75.15} & \cellcolor{sb!90} \textbf{72.01} & \cellcolor{sb!90} \textbf{80.15} & \cellcolor{sb!90} \textbf{77.34} \\
\textbf{WV2} & 60.13 & \cellcolor{sb!30} 58.87 & \cellcolor{sb!30} 63.71 & \cellcolor{sb!30} 60.90 & \cellcolor{sb!30} 71.03 & \cellcolor{sb!30} 69.72 & \cellcolor{sb!30} 76.03 & \cellcolor{sb!30} 73.84 \\
\textbf{MS}  & 59.91 & 58.24 & 63.64 & 60.28 & 68.72 & 65.23 & 74.56 & 71.00 \\
\textbf{XS}  & \cellcolor{sb!30} 60.43 & 58.73 & 62.11 & 60.29 & 69.81 & 65.63 & 75.49 & 73.26 \\
\midrule
\multicolumn{9}{c}{\textbf{TEXT}} \\ 
\midrule
\textbf{BT} & \cellcolor{sb!56} 68.74 & \cellcolor{sb!56} 62.68 & \cellcolor{sb!56} 71.54 & \cellcolor{sb!56} 67.22 & 75.48 & 73.19 & 81.85 & 79.36 \\
\textbf{XL} & 65.41 & 60.56 & 67.69 & 63.99 & \cellcolor{sb!90} \textbf{80.57} & \cellcolor{sb!56} 77.21 & \cellcolor{sb!56} 83.37 & \cellcolor{sb!90} \textbf{81.64} \\
\textbf{E5} & \cellcolor{sb!30} 66.85 & \cellcolor{sb!90} \textbf{63.72} & \cellcolor{sb!30} 70.29 & \cellcolor{sb!30} 65.86 & \cellcolor{sb!56} 80.18 & \cellcolor{sb!90} \textbf{78.91} & \cellcolor{sb!30} 82.06 & \cellcolor{sb!30} 80.43 \\
\textbf{Q3} & \cellcolor{sb!90} \textbf{70.97} & \cellcolor{sb!30} 62.06 & \cellcolor{sb!90} \textbf{72.43} & \cellcolor{sb!90} \textbf{69.14} & \cellcolor{sb!30} 79.23 & \cellcolor{sb!30} 76.79 & \cellcolor{sb!90} \textbf{84.06} & \cellcolor{sb!56} 81.48 \\
\midrule\multicolumn{9}{c}{\textbf{AUDIO+TEXT}} \\
\midrule
\multirow{3}{*}{\textbf{Concat}} 
 & \multicolumn{4}{c|}{\textbf{LTLO}} 
 & \multicolumn{4}{c}{\textbf{LOLO}} \\
\cmidrule(lr){2-5} \cmidrule(lr){6-9}
& \multicolumn{2}{c}{\textbf{ORBIT}(\ding{55}~CA)}  & \multicolumn{2}{c|}{\textbf{ORBIT}(\ding{51}~CA)}  & \multicolumn{2}{c}{\textbf{ORBIT}(\ding{55}~CA)} & \multicolumn{2}{c}{\textbf{ORBIT}(\ding{51}~CA)} \\
\cmidrule(lr){2-3} \cmidrule(lr){4-5} \cmidrule(lr){6-7} \cmidrule(lr){8-9}
 & \textbf{A} & \textbf{F1} & \textbf{A} & \textbf{F1} 
 & \textbf{A} & \textbf{F1} & \textbf{A} & \textbf{F1} \\
\midrule
\textbf{mH + BT} & 78.37 & 76.62 & 80.90 & 79.28 & 82.36 & 81.07 & 84.14 & 83.09 \\
\textbf{mH + XL} & 82.41 & \cellcolor{sb!30} 80.95 & 83.78 & 81.96 & 84.54 & 83.19 & 85.29 & 84.01 \\
\textbf{mH + E5} & 81.59 & 80.18 & \cellcolor{sb!90}\textbf{85.49} & 82.13 & 83.69 & 82.37 & 84.56 & 83.67 \\
\textbf{mH + Q3} & 80.63 & 79.69 & 83.91 & 81.70 & 84.82 & 83.22 & \cellcolor{sb!90}\textbf{86.98} & \cellcolor{sb!56} 85.29 \\
\textbf{WP + BT} & 76.56 & 75.84 & 79.43 & 77.09 & 82.77 & 81.26 & 84.09 & 83.16 \\
\textbf{WP + XL} & 81.39 & 79.61 & 82.76 & 81.19 & 83.87 & 82.37 & 84.94 & 83.06 \\
\textbf{WP + E5} & \cellcolor{sb!56} 82.65 & \cellcolor{sb!56} 81.03 & \cellcolor{sb!56} 84.97 & \cellcolor{sb!90}\textbf{83.34} & 83.12 & 82.49 & 85.41 & 83.86 \\
\textbf{WP + Q3} & 80.97 & 78.42 & 83.89 & \cellcolor{sb!30} 82.71 & \cellcolor{sb!30} 85.36 & 84.25 & \cellcolor{sb!30} 86.15 & 84.68 \\
\textbf{WV2 + BT} & 76.82 & 74.68 & 79.35 & 77.16 & 82.06 & 81.14 & 84.16 & 82.89 \\
\textbf{WV2 + XL} & \cellcolor{sb!90}\textbf{82.75} & 80.46 & 83.14 & 82.06 & \cellcolor{sb!90}\textbf{85.63} & \cellcolor{sb!90}\textbf{84.93} & \cellcolor{sb!56} 86.59 & 84.17 \\
\textbf{WV2 + E5} & 81.69 & 79.33 & 82.83 & 81.37 & 82.70 & 81.14 & 83.89 & 82.46 \\
\textbf{WV2 + Q3} & \cellcolor{sb!30} 82.48 & 80.78 & 84.67 & 80.96 & 83.42 & 82.57 & 85.17 & 84.08 \\
\textbf{MS + BT} & 75.37 & 74.51 & 79.46 & 78.69 & 85.13 & \cellcolor{sb!30} 84.77 & \cellcolor{sb!30} 85.65 & \cellcolor{sb!30} 84.90 \\
\textbf{MS + XL} & 81.90 & 80.47 & 83.97 & 82.42 & 84.91 & 83.65 & 86.01 & \cellcolor{sb!90}\textbf{85.31} \\
\textbf{MS + E5} & 82.21 & \cellcolor{sb!90}\textbf{81.39} & 83.22 & 82.28 & 83.77 & 82.99 & 85.63 & 83.29 \\
\textbf{MS + Q3} & 80.77 & 78.92 & 84.63 & \cellcolor{sb!56} 83.05 & \cellcolor{sb!56} 85.41 & 84.42 & 86.12 & 84.00 \\
\textbf{XS + BT} & 76.89 & 75.00 & 79.36 & 78.17 & 80.93 & 78.89 & 82.26 & 81.22 \\
\textbf{XS + XL} & 81.93 & 76.36 & 82.78 & 80.41 & 84.38 & 82.74 & 84.79 & 83.60 \\
\textbf{XS + E5} & 80.99 & 78.21 & 83.79 & 81.76 & 84.29 & 83.16 & 85.78 & 84.11 \\
\textbf{XS + Q3} & 81.08 & 77.98 & \cellcolor{sb!30} 84.93 & 80.11 & 85.23 & \cellcolor{sb!56} 84.85 & 85.93 & 84.21 \\
\bottomrule
\end{tabular}
\end{adjustbox}
\caption{Evaluation scores (\%) of different audio--text model combinations.}
\label{tab:2}
\vspace{-30pt}
\end{table}
Overall, these results suggest that there is no universally best unimodal PTM for cross-lingual SADD; performance is highly sensitive to language shift and modality, reinforcing the need for language-invariant multimodal fusion to achieve reliable zero-shot generalization. Therefore, by training on multiple languages, \textbf{\texttt{ORBIT}} is forced to learn language-agnostic dementia cues (rather than language-specific artifacts), which leads to consistently better zero-shot performance on held-out languages. These results demonstrate that \textbf{\texttt{ORBIT}} surpasses individual PTM baselines that previously achieved SOTA performance for SADD, including recent Whisper-based transfer-learning systems~\cite{li2024whisper,jia25_interspeech}.

\noindent \textbf{Ablation Study:} To assess the contribution of \textbf{\texttt{ORBIT}}, we conduct targeted ablation experiments. For a controlled analysis, Table~\ref{ablation44} is conducted on our best configuration, \textbf{\texttt{ORBIT}} with mH+Q3 and cross-attention. Overall, the full \textbf{\texttt{ORBIT}} model performs best, confirming that bi-geometry and language-adversarial training are crucial for strong zero-shot cross-lingual generalization.  \par

\begin{table}[!hbt]
\setlength{\tabcolsep}{13pt}
\centering
\scriptsize
\begin{adjustbox}{width=\linewidth,center}
\begin{tabular}{l|cc|cc}
\toprule
\multirow{2}{*}{\textbf{Methods}} & \multicolumn{2}{c}{\textbf{LTLO}} & \multicolumn{2}{c}{\textbf{LOLO}} \\
\cmidrule(lr){2-3}\cmidrule(lr){4-5}
 & Accuracy & F1 & Accuracy & F1 \\
\midrule
Only Hyperbolic         & 79.83 & 76.59 & 82.65 & 80.18 \\
Only Sphere             & 76.72 & 74.34 & 77.81 & 76.97 \\
w/o GRL                 & 70.99 & 68.48 & 74.53 & 71.41 \\
Hyperbolic + Euclidean  & 81.36 & 80.00 & 83.78 & 81.69 \\
Sphere + Euclidean      & 79.28 & 77.60 & 81.31 & 79.78 \\ 
\midrule
\textbf{\texttt{ORBIT}}(Ours)                  
& 83.91 & 81.70 & 86.98 & 85.29 \\
\bottomrule
\end{tabular}
\end{adjustbox}
\caption{Ablation results.}
\label{ablation44}
\vspace{-25pt}
\end{table}

\section{Conclusion}

In this study, we address zero-shot cross-lingual SADD and show that multimodal fusion is a stronger choice than relying on a single modality. Multilingual speech and text encoders capture complementary acoustic–linguistic markers of impairment, but strong transfer requires explicitly suppressing language-specific cues. Building on this insight, we propose \texttt{\textbf{ORBIT}}, a language-invariant multimodal framework that combines cross-attentive fusion, multi-tap adversarial training, and complementary spherical–hyperbolic geometric learning. Across zero-shot cross-lingual settings, \texttt{\textbf{ORBIT}} consistently outperforms unimodal baselines and simple concatenation-based fusion, providing a strong reference point for future work on cross-lingual SADD.


\section{Acknowledgements}

This work was supported by the SPEECH-D project funded by Alzheimer's Research UK (ARUK). The authors gratefully acknowledge the support of the United States--Ireland--Northern Ireland R\&D Partnership Programme (USI-207), and access to the Tier 2 High-Performance Computing resources from the Northern Ireland High Performance Computing (NI-HPC) service funded by EPSRC (EP/T022175).


\section{Use of Generative AI Disclosure}
AI assistant help was used only to polish the writing—fixing grammar, improving clarity, and making the manuscript easier to read. It did not affect the study’s ideas, analyses, results, or interpretation. The authors take full responsibility for the accuracy and integrity of the work.

\bibliographystyle{IEEEtran}
\bibliography{mybib}

\end{document}